
\input harvmac
\def\3he{{$^3${\rm He}}}

\def\ie{{\it i.e.,\ }}
\def\etal{{\it et al.}}

\def\slD{\raise.15ex\hbox{$/$}\kern-.53em\hbox{$D$}}
\def\dsl{\raise.15ex\hbox{$/$}\kern-.57em\hbox{$\Delta$}}
\def\slp{{\raise.15ex\hbox{$/$}\kern-.57em\hbox{$\partial$}}}
\def\nsl{\raise.15ex\hbox{$/$}\kern-.57em\hbox{$\nabla$}}
\def\sla{\raise.15ex\hbox{$/$}\kern-.57em\hbox{$\rightarrow$}}
\def\slla{\raise.15ex\hbox{$/$}\kern-.57em\hbox{$\lambda$}}
\def\slb{\raise.15ex\hbox{$/$}\kern-.57em\hbox{$b$}}
\def\lnp{\raise.15ex\hbox{$/$}\kern-.57em\hbox{$p$}}
\def\lnk{\raise.15ex\hbox{$/$}\kern-.57em\hbox{$k$}}
\def\lnK{\raise.15ex\hbox{$/$}\kern-.57em\hbox{$K$}}
\def\lnq{\raise.15ex\hbox{$/$}\kern-.57em\hbox{$q$}}

\def\lbar{{\mathchar'26\mskip-9mu\lambda}}

\def\cL{{\cal L}}
\def\cM{{\cal M}}


\def\pmb#1{\setbox0=\hbox{$#1$}%
\kern-.025em\copy0\kern-\wd0
\kern.05em\copy0\kern-\wd0
\kern-.025em\raise.0433em\box0 }

\def\q2{{Q^2}}
\def\gtwid{\raise.3ex\hbox{$>$\kern-.75em\lower1ex\hbox{$\sim$}}}
\def\ltwid{\raise.3ex\hbox{$<$\kern-.75em\lower1ex\hbox{$\sim$}}}
\def\12{{1\over2}}
\def\part{\partial}

\def\low#1{\lower.5ex\hbox{${}_#1$}}

\def\psl{\raise.15ex\hbox{$/$}\kern-.57em\hbox{$\partial$}}
\def\partt{\raise.15ex\hbox{$\widetilde$}{\kern-.37em\hbox{$\partial$}}}

\def\topppageno1{\global\footline={\hfil}\global\headline
={\ifnum\pageno<\firstpageno{\hfil}\else{\hss\twelverm --\ \folio
\ --\hss}\fi}}

\def\toppageno2{\global\footline={\hfil}\global\headline
={\ifnum\pageno<\firstpageno{\hfil}\else{\rightline{\hfill\hfill
\twelverm \ \folio
\ \hss}}\fi}}

\def\prl#1{Phys.\ Rev.\ Lett.\ {\bf #1}}

\def\ie{{\it i.e.},\ }

\def\nsection#1 #2{\leftline{\rlap{#1}\indent\relax #2}}

\def\prl#1{Phys.\ Rev.\ Lett.\ {\bf #1}}


%
%
%
%
\ifx\answ\bigans
\else
\output={
  \almostshipout{\leftline{\vbox{\pagebody\makefootline}}}\advancepageno
}
\fi
%
%
%
\def\washu{\vbox{\sl\centerline{Department of Physics CB1105}%
\centerline{Washington University }
\centerline{ St. Louis, MO 63130-4899, USA}}}
%
%
\def\doe{\#DOE-2FG02-91ER40628}
%
%
\def\WASHU#1#2{\noindent#1\hfill #2%
\bigskip\supereject\global\hsize=\hsbody%
\footline={\hss\tenrm\folio\hss}}
%
%
\def\abstract#1{\centerline{\bf Abstract}\nobreak\medskip\nobreak\par #1}
%
%
%
%
\edef\tfontsize{ scaled\magstep3}
 \tfontsize  \tfontsize
\font\titlermss=cmr5 \tfontsize \font\titlei=cmmi10 \tfontsize
\font\titleis=cmmi7 \tfontsize \font\titleiss=cmmi5 \tfontsize
\font\titlesy=cmsy10 \tfontsize \font\titlesys=cmsy7 \tfontsize
\font\titlesyss=cmsy5 \tfontsize  \tfontsize
\skewchar\titlei='177 \skewchar\titleis='177 \skewchar\titleiss='177
\skewchar\titlesy='60 \skewchar\titlesys='60 \skewchar\titlesyss='60
%
%
%
\def\sun{$SU(n)$}
\def\sunk{$SU(n|k)$}
\def\sunktwo{$SU(2n|2k)$}
\def\sunmk{$SU(n-k)$}
\def\sep{super-$\eta'$}
\def\etap{$\eta'$}

\def\etapnmk{$\eta'_{SU(n-k)}$}
\def\etapn{$\eta'_{SU(n)}$}
%
%
%
\def\inv{^{\raise.15ex\hbox{${\scriptscriptstyle -}$}\kern-.05em 1}}
\def\lbar{{\lower.35ex\hbox{$\mathchar'26$}\mkern-10mu\lambda}} 

%
%
%
%
\def\dsl{\,\raise.15ex\hbox{/}\mkern-13.5mu D} 
\def\delsl{\raise.15ex\hbox{/}\kern-.57em\partial}
\def\Ksl{\hbox{/\kern-.6000em\rm K}}
\def\Asl{\hbox{/\kern-.6500em \rm A}}
\def\Dsl{\hbox{/\kern-.6000em\rm D}} 
\def\Qsl{\hbox{/\kern-.6000em\rm Q}}
\def\gradsl{\hbox{/\kern-.6500em$\nabla$}}
%
%
\def\lspace{\ifx\answ\bigans{}\else\qquad\fi}
\def\lbspace{\ifx\answ\bigans{}\else\hskip-.2in\fi} 
%
%
\def\boxeqn#1{\vcenter{\vbox{\hrule\hbox{\vrule\kern3pt\vbox{\kern3pt
        \hbox{${\displaystyle #1}$}\kern3pt}\kern3pt\vrule}\hrule}}}
%
%
\def\mbox#1#2{\vcenter{\hrule \hbox{\vrule height#2in
\kern#1in \vrule} \hrule}}
%
%
%
%
\def\CA{{\cal A}}   \def\CD{{\cal D}}

 \def\CV{{\cal V}}  
 
%
%
%
%
%

%

\def\bar#1{\overline{#1}}

\def\darr#1{\raise1.5ex\hbox{$\leftrightarrow$}\mkern-16.5mu #1}

%
%
\def\frac#1#2{{\textstyle{#1\over #2}}} 
\def\bigfrac#1#2{{{#1\over #2}}} 
%
%
%
%
\def\tr{\mathop{\rm tr}}

%
%
%
%

%
%
\def\ltap{\ \raise.3ex\hbox{$<$\kern-.75em\lower1ex\hbox{$\sim$}}\ }
\def\gtap{\ \raise.3ex\hbox{$>$\kern-.75em\lower1ex\hbox{$\sim$}}\ }
\def\gl{\ \raise.5ex\hbox{$>$}\kern-.8em\lower.5ex\hbox{$<$}\ }
\def\roughly#1{\raise.3ex\hbox{$#1$\kern-.75em\lower1ex\hbox{$\sim$}}}
%
%
\def\ie{\hbox{\it i.e.}}        
        \def\cf{\hbox{\it cf.}}
\def\etal{\hbox{\it et al.}}

\def\np#1#2#3{{Nucl. Phys. } B{#1} (#2) #3}
\def\pl#1#2#3{{Phys. Lett. } {#1}B (#2) #3}
\def\prl#1#2#3{{Phys. Rev. Lett. } {#1} (#2) #3}
\def\physrev#1#2#3{{Phys. Rev. } {#1} (#2) #3}

\relax

\def\lta{\ \hbox{\raise.55ex\hbox{$<$}} \!\!\!\!\!
\hbox{\raise-.5ex\hbox{$\sim$}}\ }
\def\gta{\ \hbox{\raise.55ex\hbox{$>$}} \!\!\!\!\!
\hbox{\raise-.5ex\hbox{$\sim$}}\ }


\def\np#1#2#3{Nucl. Phys. {\bf B#1}, #3 (#2)}
\def\pl#1#2#3{Phys. Lett. {\bf B#1}, #3 (#2)}
\def\physrev#1#2#3{Phys. Rev. {\bf #1}, #3 (#2)}
\def\prl#1#2#3{Phys. Rev. Lett. {\bf #1}, #3 (#2)}

\def\CAsl{\, \hbox{/\kern-.7000em $\CA$}\, }
\def\CVsl{\hbox{/\kern-.6500em $\CV$}}
\def\CDsl{\hbox{{\ {/\kern-.6500em $\CD$}}}}
\def\frac#1#2{{\textstyle{#1 \over #2}}}

\def\gamf{\gamma_5}

\def\({\left(}\def\){\right)}

\def\[{\left[}
\def\]{\right]}
\def\({\left(}
\def\){\right)}


\nref\morel{A.~Morel, J.~Physique {\bf 48},111(1987).}
\nref\sharpeone{S.R.~Sharpe,
\physrev{D41}{1990}{3233}; Nucl. Phys. {\bf B}(Proc.Suppl.) {\bf 17},
 146 (1990);
G.~Kilcup \etal, \prl{64}{1990}{25}; S.R.~Sharpe, DOE/ER/40614-5, to be
published in {\it Standard Model, Hadron Phenomenology and Weak Decays
on the Lattice}, ed. G.~Martinelli, World Scientific.}
\nref\usPRD{C.W.~Bernard and M.F.L.~Golterman,
\physrev{D46}{1992}{853};
 Nucl. Phys. {\bf B}(Proc.Suppl.) {\bf 26}, 360 (1992).}
\nref\uslatams{C.W.~Bernard and M.F.L.~Golterman,
Nucl. Phys. {\bf B}(Proc.Suppl.) {\bf 30}, 217 (1993).}
\nref
\sharpetwo{S.R.~Sharpe, \physrev{D46}{1992}{3146};
Nucl. Phys. {\bf B}(Proc.Suppl.) {\bf 30}, 213 (1993).}
\nref\golsmi{M.F.L.~Golterman and J.~Smit, \np{245}{1984}{61}.}
\nref\gock{M.~G\"ockeler, \pl{142}{1984}{197}.}
\nref\shthwe{H.S.~Sharatchandra, H.J.~Thun and P.~Weisz, \np{192}{1981}{205}.}
\nref\vddsmi{C.~van~den~Doel and J.~Smit, \np{228}{1983}{122}.}
\nref\mpr{E.\ Marinari, G.\ Parisi and C.\ Rebbi, \np{190}{1981}{734}.}
\nref\smivin{J.~Smit and J.~Vink, \np{286}{1987}{485}.}
\nref\golsmilet{M.F.L.~Golterman and J.~Smit, \pl{140}{1984}{392}.}

\vskip 1.2in
\centerline{{\titlefont{Partially Quenched Gauge Theories and}}}
\vskip .2in
\centerline{{\titlefont{ an Application to Staggered Fermions}}}
\bigskip
\centerline{{\it Submitted to Physical Review D}}
\vskip .5in
\centerline{ Claude\ W.\  Bernard and Maarten\ F.L.\ Golterman}
\bigskip
\washu
\bigskip
\bigskip

\abstract{
We extend our lagrangian technique for chiral perturbation theory for
quenched QCD to include theories in which only some of the quarks are quenched.
We discuss the relationship between the partially quenched theory and a
theory in which only the unquenched quarks are present.  We also investigate
the peculiar infrared divergences associated with the $\eta'$ in the quenched
approximation, and find the conditions under which such divergences can
appear in a partially quenched theory.
We then apply our results to staggered fermion QCD
in which the square root of the fermion determinant is taken, using the
observation
that this should correspond to a theory with four quarks, two of which
are quenched.}
\vskip 2.0truein
\line{e-mail: cb@lump.wustl.edu, maarten@wuphys.wustl.edu\hfill}
\WASHU{\hbox{Wash.~U.~HEP/93-32}}{May, 1993}

\eject

\newsec{Introduction}

There has been a growing interest in chiral perturbation theory
(ChPT)
for the quenched approximation of QCD \refs{\morel-\sharpetwo}.
The motivation for
this is the fact that at present
the quenched approximation is indispensable for
the numerical study of QCD.  Since ChPT is
used in order to analyze and extrapolate numerical data, it is necessary
to adapt it to the quenched approximation.  To this end, we have
developed a systematic, lagrangian technique that can be used
straightforwardly to calculate quenched correlation functions within
ChPT \usPRD.  Quenched  ChPT has been applied to calculate several
important quantities, such as masses and decay constants
\refs{\morel-\sharpetwo}.

However, aside from these practical results, it turns out that the
special role of the $\eta'$ in the quenched approximation leads to a
very singular infrared behavior of quenched ChPT, which suggests that
quenched QCD does not have a chiral limit \refs{\usPRD-\sharpetwo}.
Unlike the
case in the full theory, the mass of the $\eta'$ cannot be taken to
infinity so as to leave us with an effective action which describes only
the pseudo-Goldstone mesons.  Instead, a double pole term
proportional to the singlet part of the $\eta'$ mass squared shows up in the
$\eta'$ twopoint function and  causes the peculiar infrared behavior.

In this paper, we wish to extend our investigations to the case of
partially quenched theories.  Partially quenched theories are theories
in which not all fermions are quenched; only for some of the fermions
present in the theory will the determinant in the functional integral
be replaced by $1$.
(We assume throughout a bilinear fermion action
with only flavor diagonal terms.)  The same method that we developed for
studying ChPT for a completely quenched theory can also be applied in
this case.  Our motivation for considering partially quenched theories
is threefold:

First, one may learn more about the peculiar infrared
behavior by considering what happens when only part of the fermion
content of a theory is quenched.
In particular, if different fermion mass scales
are present, one might ask how the
infrared behavior depends on whether all or only some of the fermions
with a common mass are quenched.  Also, it is interesting to know what
happens in the unquenched sector of the theory: is a
theory with $n$ fermions, out of which $k$ are quenched, the same as an
unquenched theory with just $n-k$ fermions?  In the first three sections
of this paper we address these questions.

Second,
partially quenched theories arise naturally in the
description of simulations in which the valence quark masses are not
chosen  equal to the sea-quark masses.  This is a not uncommon
numerical technique which, for example, allows one to use
Wilson valence quarks and staggered  sea-quarks.  One would
like to have a chiral theory for such simulations.

A third motivation comes from staggered fermions themselves.
It is well
known that lattice QCD with staggered fermions describes QCD with four
flavors of quarks in the continuum limit.  In order to use these
fermions for simulations of QCD with only two flavors, a trick which has
been used is to take the square root of the fermion
determinant, thereby effectively reducing the number of flavors which
appear in virtual quark loops from four to two.

This approach seems justified in weak coupling perturbation theory, but
of course the question is whether it is really a legitimate technique.
Certainly, taking the square root is not equivalent to formulating a two
flavor theory through a functional integral with a local lagrangian.
In the continuum limit, however, taking the square root of a four flavor
fermion
determinant (with at least pairwise degenerate quark masses)
is exactly equivalent to quenching two out of four flavors, and our
results about partially quenched ChPT with $n=4$ and $k=2$ should apply.
This then allows us to test the idea of taking the square root within the
context of ChPT:
if this trick is legitimate, and does indeed lead to a two flavor theory, that
should be reflected in ChPT.  It means that partially quenched ChPT
should reproduce the results of unquenched ChPT as long as we allow only
unquenched quarks on the external lines of correlation functions
(including operators which excite bound states of unquenched quarks).

Using results obtained in the first part of the paper, we show that this is
indeed the case.  This is a nontrivial test
of the trick of taking the square root and
complements the argument based on weak coupling perturbation theory, since
ChPT addresses a different regime of QCD.  As a corollary, we present a
simple way in which $\eta'_{SU(2)}$ correlation functions can be
computed in numerical simulations.

\newsec{Theorems}

In this section we state three theorems about partially quenched
gauge theories and then give the physical arguments which
underlie these results.  We leave to sections 3 and 4 the detailed
calculations  in quenched chiral perturbation theory
\refs{\usPRD}
which illustrate the theorems and various corollaries.
For the first two theorems,
the physical arguments actually constitute   proofs, and the calculations
of section 4 serve as explicit examples. For the third theorem, however,
we give only an argument here only for a special case; the full proof
will have to wait until section 3.

We first need to establish our notation.   Consider a QCD-like
theory with $n$ flavors of quarks, $q_i$.  The  quark masses
$m_i$ ($i = 1, \dots n$) are completely arbitrary.  We
then partially quench this theory by adding $k$ flavors ($0 \le k \le n$)
of pseudoquarks (bosonic quarks), $\tilde q_j$, as in ref.\ \usPRD.
Note that the limiting
cases of complete quenching or no quenching  are allowed.
The masses of the pseudoquarks are fixed to be equal to the masses
of the first $k$ real quarks, \ie,
$m_j$ ($j = 1, \dots k$).  In other words the first
$k$ quarks are ``quenched,'' and the remaining $n-k$ quarks
are ``unquenched.''
We call this theory
the ``\sunk\  theory.''
We do not consider
the more general, but probably physically uninteresting, case
where the masses of the
pseudoquarks are arbitrary ---
in that case there can be virtual pseudoquark loops that do not cancel
completely against real quark loops.
Note that,
if there are degeneracies between some unquenched and
some quenched quarks, \ie, the quarks of some mass scale are
only partially quenched, one is free to choose which of the
quarks shall be the considered the ``unquenched'' ones.  In other
words, if among $a$ degenerate quarks, $b$ ($b<a$) are quenched,
one may arbitrarily choose which $b$ quarks
are to have indices $j$, with $1\le j\le k$, and which $b-a$ quarks are to have
indices
$r$, with $k+1 \le r \le n-k$.

A normal, completely unquenched
theory with $n$ quarks will be denoted as an ``$SU(n)$ theory;'' it is
obviously the same as the $SU(n|0)$ theory.

The full chiral symmetry of the $SU(n|k)$ theory is
the semi-direct product\foot{The product is obviously
direct when $k=0$.}
$[SU(n|k)_L\otimes SU(n|k)_R]{\bigcirc\kern -0.25cm s\;}
U(1)$, where the additional $U(1)$ present in
$U(n|k)_L\otimes U(n|k)_R$ is broken by the anomaly.

The special case where all the $m_i$ ($i = 1, \dots n$) masses
are equal is called the ``degenerate \sunk\ theory.''
In section 4, we also examine
another special case where the number of
quarks and pseudoquarks are even ($n\to2n$, $k\to2k$) and the
masses just take on two values: $m_1 = m_3 = m_5 = \cdots = m_{2n-1}
\equiv m_u$,
and $m_2 = m_4 = m_6 = \cdots = m_{2n} \equiv m_d$.  We call
this the ``doublet \sunktwo'' theory.

We define the ``\sep'' of the \sunk\ theory by the
interpolating field
\eqn\phizero{\Phi_0 = c(\sum_{i=1}^n \bar q_i \gamma_5 q_i +
\sum_{j=1}^k \bar{\tilde q}_j \gamma_5 \tilde q_j)\ .}
The normalization factor $c$ was taken to
be $1/\sqrt{n+k}$\ \  (\ie, $1/\sqrt{6}$ for $n=k=3$) in
ref.\ \usPRD, but has been left
arbitrary here, since different normalizations will
often be useful.  The reader may be confused by the fact the
\sep\ does not appear explicitly as the {\sl difference}
of the quark \etap\ and the pseudoquark \etap, as
it does in the corresponding chiral theory, but rather as their
sum.  The reason is that the mesonic fields of a chiral
theory directly correspond not to $\bar q \gamma_5q$, but
to
$tr(q \bar q \gamma_5)$, since it is the {\sl first}
index of the meson matrix $\Sigma$ which is the quark index.
Taking into account the opposite statistics
of quarks and pseudoquarks, the relative minus sign between
quark and pseudoquark \etap\ would reappear in eq.\ \phizero\ when
written in the latter form.

We can now state the three theorems about partially
quenched theories:

\item{I.} In the subsector where all valence quarks
are unquenched (\ie, where all valence quarks are of type
$r$, where $k+1 \le r \le n)$, the \sunk\ theory is completely
equivalent to a normal, completely unquenched \sunmk\ theory.

\item{II.} The \sep\ (with normalization $c = 1/\sqrt{n-k}$
in eq.\ \phizero) is equivalent to the \etap\ constructed in
the unquenched sector of the \sunk\ theory, and is therefore,
by I, equivalent to the \sunmk\ \etap.  ``Equivalent''
here means that Green's functions constructed from an arbitrary
number of \sep\ fields and unquenched quarks, will be equal to
the corresponding Green's functions with the \sep\ replaced
by the \etap\ of the unquenched sector of the
\sunk\ theory (or, what is the same, by the \sunmk\ \etap).
Green's functions which
involve the \sep\ and arbitrary combinations of quenched
quarks or pseudoquarks are not allowed --- there is nothing
for them to correspond to in the \sunmk\ theory.

\item{III.} Quenched infrared divergences\refs{\usPRD-\sharpetwo},
 coming from a double
pole in the \etap\ propagator and associated with some quark
mass of mass $m_j$, will arise if and only if the scale
$m_j$ is fully quenched, \ie, if there is a pseudoquark of mass $m_j$
for every quark of mass $m_j$.  In other words, these
unphysical divergences arise if and only if $1\le j \le k$
(so that this quark is quenched)
and $m_j \not= m_r$, for all $r$ with $ k+1 \le r \le n$
(so that there are no unquenched quarks of the same mass).

Theorem I is easily established by a simple argument.  Since by supposition
all the valence quarks are unquenched, the only way the
amplitudes
could ``know'' about the quenched quarks and pseudoquarks
is through virtual loops.  But the pseudoquarks have been chosen to
cancel the quenched quarks exactly in virtual loops,
so only the unquenched quarks can appear anywhere in a diagram.

Theorem II also relies on the cancellation between quenched
quarks and the pseudoquarks, but this time in valence
lines.  There are two quark flow diagrams that contribute
to the \etap\ propagator: the ``straight-through''
diagram (Fig.~1a) and the ``two-hairpin'' diagram
(Fig.~1b). Note that we only specify the valence quark
lines in these diagrams; in general (for $n\not=k$) there will
be additional virtual quark loops.  By the definition
of the \sep\  and the opposite statistics of quarks and
pseudoquarks,
the pseudoquarks will cancel the quenched quarks
both in the straight-through and the two-hairpin
diagrams (in the latter case, the cancellation takes
place separately in each hairpin).  Only the unquenched
quarks survive in each diagram, and they of course
will also be the only survivors in any virtual quark
loops.  Thus the  contraction of any two \sep\
fields is the same as the contraction of two \sunmk\
\etap\ fields.  (The choice $c =  1/\sqrt{n-k} $ reproduces
the canonical normalization.)  Similarly, the
contraction of a \sep\ with some combination of unquenched
quark fields is the same as the contraction of an
\sunmk\ \etap\ with the same combination, since only
the unquenched quarks in the \sep\ will contribute.

\topinsert \vskip 5.0truecm
\baselineskip=12truept\parindent=0pt

{\titlermss  [Instructions for do-it-yourself graphics:
\medskip

For Fig.~1(a), draw an oval (racetrack) with its long axis horizontal.
Put an ``x'' on  each short end (\ie, the extreme right and left edges).
\medskip

For Fig.~1b), draw an two ovals as in (a) and put them end to end.
Put an ``x'' on the left edge of of the left oval and on the right
edge of the right oval.
\medskip

Good going, you did it!]}

\bigskip

\qquad\qquad\qquad\qquad\qquad (a) \qquad\qquad\qquad\qquad\qquad\qquad\qquad
(b)

\centerline{Figure 1}

{\it Quark flow diagrams for the $\eta'$ propagator;
(a) is the ``straight-through'' diagram, and (b) is the
``two-hairpin'' diagram.  Arbitrary numbers of gluon corrections
and virtual quark loops (if the theory is not fully quenched)
are implicit.}
\endinsert

Theorem III relies on the fact that the infrared divergences
that have been found \refs{\usPRD-\sharpetwo} in the quenched
theory arise from the two-hairpin diagram (Fig.\ 1b),
which gives a double pole in the \etap\ propagator.
In section 3, we calculate the propagator in the neutral
meson sector in partially quenched chiral perturbation theory,
and show that the offending double poles can only arise when
a mass scale is fully quenched.  Here, we make this result
plausible by examining the degenerate
\sunk\ theory.  It is not difficult to argue that, when
there is only one quark mass scale, double
poles arise only when the theory is completely quenched,
\ie, when $k=n$.

Consider the propagator of the  \etap\ of the  degenerate \sunk\ theory
(constructed from the
$n$ quarks only, with no pseudoquarks).
In a normal unquenched theory, the two-hairpin contribution
to the propagator would
include diagrams with arbitrary numbers of
virtual quark bubbles
between the two hairpins.  The set of all
these diagrams, together with the
straight-through diagram, is a geometric series
which can be summed to a simple pole, with an \etap\ mass
shifted away from the common meson multiplet mass by
the usual
singlet contribution.  In a partially
quenched theory, the sum over bubbles is still present, and differs
only by an overall normalization (coming from the counting of the
{\sl valence} loops in the hairpin) relative to
the case where only the
unquenched quarks are allowed.  Therefore, the two-hairpin diagram
has an ``incorrect'' normalization relative to the
straight-through diagram, so the full propagator will
not just be a simple pole with a singlet mass term added.
We can correct for this mismatch by adding and subtracting
the proper amount of straight-through diagram.
The complete \etap\ propagator will then be a sum of two
simple poles, one with mass equal to the common meson
multiplet mass (from the subtracted
piece of the straight-through diagram)  and one with a shifted mass
which includes
the singlet contribution.  Since there are no double poles,
there will be no unusual infrared divergences.  The only exception
occurs for $k=n$, when there are no bubbles to sum.
Then the complete \etap\ propagator is just
the sum of two terms:  a double pole from two-hairpin diagram
and a single pole from the straight-through diagram \refs{\usPRD,\uslatams}.

An example which illustrates several of the ideas discussed
above is the construction, in a degenerate
\sunk\ theory, of an \sunmk\ \etap\
out of the diagrams for the \sun\ \etap.
Let both of these particles
be described by canonically normalized fields:

\eqn\etapnorm{
\eqalign{\eta'_{SU(n-k)} =& {1\over \sqrt{n-k}}
                  \sum_{i=1}^{n-k} \bar q_i \gamma_5 q_i, \cr
         \eta'_{SU(n)} =& {1\over \sqrt{n}}
                  \sum_{i=1}^n \bar q_i \gamma_5 q_i \ . \cr}}

The  straight-through diagram of Fig.~1a is then clearly
identical for the \etapn\ and the \etapnmk, since the
factor of $n$ or $n-k$ from flavor counting in the loop
cancels against the field normalization
factors.
The two-hairpin diagram (Fig.~1b) is however
normalized differently for the \etapn\ and the \etapnmk,
since there are now two flavor loops.  To get the
\etapnmk\ two-hairpin from the \etapn\ two-hairpin,
one must multiply the latter by $(n-k)/n$.  This
holds irrespective of the number of virtual loops in the two-hairpin
diagram.
Note that each virtual bubble
is normalized the same in both cases,
since we are always working in a \sunk\ theory with
a net total of $n-k$ flavors in virtual loops.  The difference
in normalization arises only from the valence (hairpin) loops.

We may thus obtain the correct \etapnmk\ propagator
in an \sunk\ theory from the \etapn\ propagator
in that same theory by making a simple readjustment
of the relative weights of the diagrams. This will
prove useful when we discuss staggered fermions.

\newsec{Proof of Theorem III}

We begin by writing down the lagrangian for the \sunk\ theory.
Define the $(n+k)\times(n+k)$ hermitian field
 $\Phi$ by
\eqn\phidef{\Phi\equiv\left(\matrix{\phi&\chi^\dagger\cr
                          \chi&\tilde \phi\cr}\right),}
where $\phi$ is the $n\times n$ matrix of ordinary mesons
made from the $n$ ordinary quarks and their
antiquarks, $\tilde \phi$ is
the corresponding $k\times k$ matrix for pseudoquark mesons,
and $\chi$ is a $k \times n$ matrix of mesons made from
a pseudoquark and an ordinary antiquark.
The unitary field $\Sigma$ is then defined as
\eqn\sigmadef{\Sigma \equiv \exp(2 i \Phi/f)\ ,}
with $f$ the tree-level pion decay constant.
 The $(n+k)\times(n+k)$ quark mass matrix is given by
\eqn\masses{\cM_{ij} = m_i \delta_{ij}\ ,}
where, as discussed in the
previous section, the masses $m_i$ for $i=1, \dots, n$ are
arbitrary, and we take the pseudoquark masses
equal to the first $k$ quark masses:
$m_{n+j}\equiv m_j$ for  $j=1, \dots, k$.

The euclidean lagrangian is then
\eqn\lagrangian{\eqalign{
\cL =&
V_1(\Phi_0)str
(\partial_\mu\Sigma\partial^\mu\Sigma^\dagger)
-V_2(\Phi_0)str(\cM\Sigma+\cM\Sigma^\dagger)\cr
&\phantom{junk}+ V_0(\Phi_0)+V_5(\Phi_0)(\partial_\mu\Phi_0)^2,}}
where the functions $V_i$ can be chosen
to be real and even by making use
of the freedom allowed by field redefinitions \usPRD.
We choose $\Phi_0=str(\Phi)$ which corresponds to
$c=1$ in eq.\ \phizero. Since the
two-hairpin-like interactions between neutral
mesons must have no dependence on $n$ or $k$ at tree level, this choice of $c$
guarantees that the parameters in the expansion
of the $V_i$ are $n$- and $k$-independent at tree level.

For the purposes of this section we just need the quadratic
terms in \lagrangian.  We define
\eqn\mudef{\eqalign{
V_1(0)& \equiv {f^2 \over 8}\ ,\cr
V_2(0)& \equiv v \left(= {f^2 m_{\pi^+}^2 \over 4(m_u + m_d)}\right)\ ,\cr
V''_0(0)& \equiv {\mu^2 \over 3}\ ,\cr
V_5(0)& \equiv {\alpha \over 6}\ .}}
Note that $V_5(0)$ and $V''_0(0)$ are not the
same as in ref.\ \usPRD\ because of the different
choice of normalization for $\Phi_0$.

We are now in the position to prove theorem III.
In the case of arbitrary quark mass the simplest
approach is just to calculate the
neutral-meson propagator explicitly in tree approximation.
We work in the basis of the states $U_i$, $i=1,\dots,n+k$
corresponding to $\bar u u,
\bar d d, \bar s s, \dots$, and their pseudoquark counterparts.  From
eqs.\lagrangian\ and \mudef, the neutral
inverse propagator in momentum space is
\eqn\ginverse{G^{-1}_{ij} = \delta_{ij}(p^2 + M_i^2)\epsilon_i + {\mu^2 \over
3}
\epsilon_i\epsilon_j,}
 where $M_i^2 \equiv 8 v m_i/f^2$,
we have taken $\alpha=0$ (it is easy to reinstate later on
by the substitution $\mu^2 \to \mu^2 + \alpha p^2$), and
$\epsilon_i$ is defined by
\eqn\epsilondef{\epsilon_i = \Bigl\{\matrix{ +1,&{\rm for\ } 1\le i\le n\cr
                                     -1,&{\rm for\ } n+1\le i\le n+k}\ \ .}

It is straightforward to invert \ginverse, either by
expanding in powers of $\mu^2$ or by guessing the
form of the inverse and fixing the coefficients
by $GG^{-1}=1$.  We have
\eqn\g{G_{ij} = {\delta_{ij} \epsilon_i \over p^2 + M_i^2} -
{\mu^2/3 \over (p^2 +M_i^2)(p^2+M_j^2) F(p^2)}\ ,}
where
\eqn\fdef{F(p^2) \equiv 1 + {\mu^2 \over 3} \sum_{i=1}^{n+k}
{\epsilon_a \over p^2 + M_i^2}\ =
 1 + {\mu^2 \over 3} \sum_{r=k+1}^{n}
{1 \over p^2 + M_r^2}\  .}
The last equality in \fdef\ follows from the fact that
the pseudoquark masses have been chosen equal to the
first $k$ quark masses.

Theorem III now follows by examination of eq.\ \g.
We show in the Appendix that $F(p^2)$ has no double
zeros, so no double poles in $G$ can arise from $F$.
Therefore, the only way there can be a double pole
in $G$ is for $M_i = M_j$, for some $i,j$ (this is of
course trivially satisfied for $i=j$), and $M_j \ne M_r$,
for all $j$ between $k+1$ and $n$.  Since $M_r$, $k+1\le r\le n$,
are just the masses of the neutral mesons composed of
unquenched quarks, the latter condition implies that
double poles occur at mass $M_j$ if and only if quarks
of the corresponding mass are completely quenched.
This is just the content of theorem III.

It is instructive to examine eq.\ \g\ in the degenerate
limit ($M_i \equiv M$ for all $i$).  For $k\not= n$, we have
\eqn\gdeg{G_{ij} = {\delta_{ij} \epsilon_i -1/(n-k) \over p^2 + M^2} +
{1/(n-k) \over p^2 +M^2 + (n-k)\mu^2/3}\ .}
This clearly illustrates a result of section 2: that the degenerate
propagator (for $k\not= n$) is the sum of two simple poles,
one with mass equal to the common meson
multiplet mass
and one with a shifted mass
which includes
the singlet contribution.   Note that for $k=n$ one sees immediately
from eqs.\ \g\ and \fdef\ that there are always double
poles in $G$ for $i=j$.

\newsec{Examples}

We would like to demonstrate the theorems in some explicit examples.
First, we will consider the case of completely degenerate quark masses,
with $n$ normal quarks and $k$ pseudoquarks.  We have calculated the
self-energies of the pion and the super-$\eta'$ to one loop.
The euclidean-space pion self-energy is

\eqn\pionse{
\eqalign{
\Sigma_\pi(p)=&-\bigfrac{2}{3}(n-k)\left(p^2+m_\pi^2\right)I(m_\pi^2)
+\bigfrac{2}{(n-k)}m_\pi^2\left(I(m_\pi^2)-
\bigfrac{I(m_{\eta'}^2)}{1+\frac{1}{3}\alpha(n-k)}\right)\cr
&+4(n-k)\left(V_1''(0)p^2+
\bigfrac{f^2V_2''(0)}{8v}m_\pi^2\right)
\bigfrac{I(m_{\eta'}^2)}{1+\frac{1}{3}\alpha(n-k)},
}
}
where $m_\pi^2=8vm/f^2$,
and

\eqn\integral{
I(m^2)=\bigfrac{1}{f^2}\int\bigfrac{{\rm d}^4q}{(2\pi)^4}\bigfrac{1}{q^2+m^2}.
}
For the purpose of this paper we do not need to specify
how we regulate such integrals.
$m_{\eta'}$ is the mass of the $\eta'$ in the $SU(n-k)$ theory, and is
given by

\eqn\etapmass{
m_{\eta'}^2=\bigfrac{\frac{1}{3}(n-k)\mu^2+m_\pi^2}{1+\frac{1}{3}\alpha(n-k)}.
}

Theorem I is clearly obeyed by this result:
the self-energy is a function of $n-k$ only, and therefore is equal to
the pion self-energy computed in a theory with $n-k$ normal quarks and no
pseudoquarks.

For $k<n$ the above results only have chiral logarithms of the standard
type, arising
from integrals over single pole propagators.
On the other hand, for $k=n$ the result is

\eqn\pionsenn{
\Sigma_{\pi,k=n}(p)=\bigfrac{2}{3}\bigfrac{m_\pi^2}{f^2}
\int\bigfrac{{\rm d}^4q}{(2\pi)^4}\bigfrac{\mu^2+\alpha q^2}{(q^2+m_\pi^2)^2}.
}
This contributes a term to the pion mass which goes like $\mu^2m_\pi^2
\log{m_\pi^2}$, unlike the usual $m_\pi^4\log{m_\pi^2}$.  It is an
example of the ``pathological terms" previously seen in quenched
calculations.\foot{In other quantities such as $\langle{\bar\psi}
\psi\rangle$ or $f_K/f_\pi$ such pathologies can lead to actual
infrared divergences as a quark mass is taken to zero, see refs.
\refs{\usPRD-\sharpetwo}.}
This is a special case of theorem III: there is a pathology as
$m_\pi\to 0$ which comes from a double pole and arises only for $k=n$,
in which case the
mass scale $m$ is fully quenched.

To demonstrate theorem II, we have to compute the self-energy for the
super-$\eta'$, $\Phi_0$.  On the external
lines, we choose $c=1/\sqrt{n-k}$ in eq.\ \phizero, so that for $k=0$
we are just calculating the conventionally
normalized \etap\ self energy in an ordinary unquenched theory.
(Recall, however, that $\Phi_0$ in the potentials $V_i(\Phi_0)$ is
normalized with $c=1$.)
For the degenerate case, the result is

\eqn\etapse{
\eqalign{
\Sigma_{\Phi_0}(p)=&\bigfrac{-2m_\pi^2}{n-k}
\left(\left[(n-k)^2-1\right]I(m_\pi^2)+
\bigfrac{I(m_{\eta'}^2)}{1+\frac{1}{3}\alpha(n-k)}\right)\cr
&+\bigfrac{f^2V_2''(0)}{2v}(n-k)m_\pi^2
\left(\left[(n-k)^2-1\right]I(m_\pi^2)+
\bigfrac{6I(m_{\eta'}^2)}{1+\frac{1}{3}\alpha(n-k)}\right)\cr
&-4V_1''(0)(n-k)
\left(\left[(n-k)^2-1\right]m_\pi^2I(m_\pi^2)
+(m_{\eta'}^2 - p^2)
\bigfrac{I(m_{\eta'}^2)}{1+\frac{1}{3}\alpha(n-k)}\right)\cr
&+f^2\left(\bigfrac{1}{2}V_0''''(0)+V_5''(0)(p^2 -m_{\eta'}^2)\right)(n-k)^2
\bigfrac{I(m_{\eta'}^2)}{1+\frac{1}{3}\alpha(n-k)}.
}
}
Again, for $k<n$ we see that this is only a function of $n-k$, and
therefore equal to the self-energy of the $\eta'$
in the $SU(n-k)$ theory.  For $k=n$ the normalization $c=1/\sqrt{n-k}$
is clearly inappropriate, and we should  multiply \etapse\ through by $(n-k)$.
The expression then vanishes when $n=k$
(note that in that case $m_\pi=m_{\eta'}$),
consistent with our expectation that the $\Phi_0$
propagator (with finite $c$) vanishes in the fully quenched theory \usPRD.

As a final example, we consider the ``doublet $SU(2n|2k)$" theory, in
which we have $n$ quarks and $k$ pseudoquarks with mass $m_u$, and $n$
quarks and $k$ pseudoquarks with mass $m_d$.  We will only present that
part of the one loop
pion self-energy which comes from the 4-meson vertex proportional to
$V_1(0)$, since the full expression is quite cumbersome.  All our
conclusions hold separately for the contribution from each vertex to
the self-energy,
since the parameters multiplying these vertices
are free.   The result is (here we set $\alpha=0$ for simplicity)

\eqn\pionsedbl{
\eqalign{
\Sigma_\pi^{V_1(0)}(p)=
&-{1\over{3f^2}}\int\bigfrac{{\rm d}^4q}{(2\pi)^4} (q^2+p^2)
\Biggl[(n-k)\left(\bigfrac{1}{q^2+m_U^2}+\bigfrac{1}{q^2+m_D^2}
+\bigfrac{2}{q^2+{\bar m}^2}\right)\cr
&-\bigfrac{1}{n-k}\left(\bigfrac{1}{q^2+m_U^2}+\bigfrac{1}{q^2+m_D^2}\right)\cr
&+\bigfrac{2}{n-k}\left(\cos^2{\theta}\bigfrac{1}{q^2+m_+^2}
+\sin^2{\theta}\bigfrac{1}{q^2+m_-^2}\right)\Biggr].
}
}
(The quartic divergence present in eq.\ \pionsedbl\ is cancelled
in the total self-energy by a
term coming from the measure of the path integral.)
In this expression,

\eqn\masses{
\eqalign{
m_U^2&=\bigfrac{8m_uv}{f^2},\cr
m_D^2&=\bigfrac{8m_dv}{f^2},\cr
{\bar m}^2&=\bigfrac{4(m_u+m_d)v}{f^2}=\bigfrac{1}{2}(m_U^2+m_D^2),\cr
m_\pm^2&=({\bar m}^2+\bigfrac{1}{3}(n-k)\mu^2)\mp
\sqrt{\bigfrac{1}{9}(n-k)^2\mu^4+\bigfrac{1}{4}(m_U^2-m_D^2)^2},
}
}
and

\eqn\sinsq{
\sin^2{\theta}=\bigfrac{n-k}{12}\bigfrac{\mu^2(m_U^2-m_D^2)^2}
{(m_-^2-m_U^2)(m_-^2-m_D^2)
\sqrt{\bigfrac{1}{9}(n-k)^2\mu^4+\bigfrac{1}{4}(m_U^2-m_D^2)^2}}.
}
The five different masses which appear in eq.\ \pionsedbl\ correspond to
the various meson masses which appear in the $SU(2(n-k))$ theory.
In the flavor off-diagonal sector, $m_U$ and $m_D$ correspond to mesons
of types ${\bar u}_iu_j$ and ${\bar d}_id_j$ with $i\ne j$, and $\bar m$
to ${\bar d}u$ or ${\bar u}d$.  In the flavor diagonal sector, there
are $2(n-k-1)$ $\pi^0$-like mesons with masses $m_U$ and $m_D$, and two
other neutral mesons with masses $m_\pm$ due to the singlet-nonsinglet mixing
which occurs for $m_u\ne m_d$.
$\theta$ is the mixing angle between these two latter neutral mesons.  In the
case that $n-k=1$, the coefficients of the $m_U^2$ and $m_D^2$ poles
in eq.\ \pionsedbl\
vanish, consistent with the meson spectrum of the $SU(2)$ theory.

For $k=n$, we have

\eqn\piosedblnn{
\Sigma_{\pi,k=n}^{V_1(0)}(p)=
\bigfrac{\mu^2}{9f^2}\int\bigfrac{{\rm d}^4q}{(2\pi)^4}(q^2+p^2)
\left(\bigfrac{1}{q^2+m_U^2}-\bigfrac{1}{q^2+m_D^2}\right)^2,
}
which is independent of $n$ since results in a fully quenched
theory must depend only on  the valence quarks.
Eq.\ \piosedblnn\
agrees for $u\to s$ with
the result we obtained in computing the
one loop corrections to the kaon mass (\cf\ ref.\ \refs{\usPRD}).

\newsec{Application to Staggered Fermions}

In this section we will apply some of the results obtained in the previous
sections to lattice QCD with staggered fermions.
In the scaling region,
this theory describes QCD with four quark flavors, which can be given
nondegenerate masses by using nonlocal mass terms \refs{\golsmi,\gock}.
If one would like to consider QCD with two
flavors, one can use the so-called reduced staggered fermion formalism
\refs{\shthwe,\vddsmi},
which however leads to a complex fermion determinant \vddsmi.
Also the reduced
staggered fermion action
does not possess any continuous chiral invariance, unlike
``normal" staggered fermions.  An alternative is to consider
normal staggered fermions and
define a two-flavor theory by taking the square root of the determinant
\refs{\mpr}.
This corresponds to quenching two of the four flavors. We therefore
expect that the low energy meson effective theory will be described by
$SU(4|2)$ chiral perturbation theory.  For this to work, the masses need to be
at least pairwise degenerate.

If only a single site mass term is used,
the staggered fermion determinant with degenerate quark masses is
positive \refs{\smivin}.
Since the continuum
limit is a degenerate four-flavor theory with a determinant which is the fourth
power of a one-flavor determinant, one expects that taking the positive
square root of the staggered fermion determinant leads to the desired
determinant for the continuum two-flavor theory.

If nonlocal mass terms
are used, the determinant is not positive in general.
However, the continuum determinant for each flavor is (formally)
positive, so one
might expect that with staggered fermions close enough to the continuum
limit, no problem arises in taking the square root.

In this section, we will consider the definition of two-flavor meson
operators in the mass degenerate two-flavor theory obtained from the
degenerate four-flavor theory in which the square root of the
determinant is taken.  Theorem I tells us that we can obtain
the two-flavor unquenched theory in this way, and that
no problems are to be expected from taking
the square root.  For nonsinglet mesons no tuning of the operators
is required because one may use the same operators as in the four-flavor
theory.

In general, however,
one will need to tune the staggered hadron operators in order
to project out the various continuum hadronic states of interest
\refs{\golsmi}.
For example, tuning will be necessary
in order to have more than one different quark mass within the
staggered fermion formalism \refs{\golsmi,\golsmilet}.
In particular, one expects that the definition of an operator for the
$\eta'_{SU(2)}$ in the four-flavor theory will require tuning, even with
degenerate quark masses.
What we wish to show here is
that nevertheless two ways exist for choosing a mass matrix and a meson
operator which do not require tuning of the operator in order to define
a pure $\eta'_{SU(2)}$ in the four-flavor theory.  The first method
consists of applying theorem II, whereas the second method makes use of
a peculiarity of nonlocal staggered fermion mass terms.

We will start by reviewing some facts about renormalization for
staggered fermions.  A general mass term for staggered fermions is given
by \refs{\golsmi}

\eqn\stagaction{
\eqalign{
S_{\rm mass}=&\sum_x m\bar{\chi}(x)\chi(x)
+\sum_{xy}m_\mu\bar{\chi}(x)E_\mu(x,y)\chi(y)
-\frac{1}{2}i\sum_{xyz}m_{\mu\nu}\bar{\chi}(x)E_\mu(x,y)E_\nu(y,
z)\chi(z)\cr
&-\frac{1}{6}i\sum_{wxyz}m^5_\mu\epsilon_{\mu\alpha\beta\gamma}
\bar{\chi}(w)E_\alpha(w,x)E_\beta(x,y)E_\gamma(y,z)\chi(z)\cr
&-\frac{1}{24}\sum_{vwxyz}m^5\epsilon_{\alpha\beta\gamma\delta}
\bar{\chi}(v)E_\alpha(v,w)E_\beta(w,x)E_\gamma(x,y)E_\delta(y,z)\chi(z).
}
}
The operator $E$ is defined through

\eqn\stagE{
\sum_{xy}m_\mu\bar{\chi}(x)E_\mu(x,y)\chi(y)=
\frac{1}{2}\sum_{z\mu}m_\mu\zeta_\mu(z)[\bar{\chi}(z)U_\mu(z)\chi(z+{\hat\mu})
+\bar{\chi}(z+{\hat\mu})U^\dagger_\mu(z)\chi(z)],
}
where the $\zeta_\mu$ are certain site-dependent sign factors (\cf\
ref.\ \refs{\golsmi}).
$m_{\mu\nu}$ is taken to be antisymmetric.

This mass term leads to the following mass
matrix $M$ for the four flavors that emerge in the continuum limit:

\eqn\stagmass{
M=m+m_\mu\xi_\mu+\frac{1}{2}m_{\mu\nu}(-i\xi_\mu\xi_\nu)+m^5_\mu
i\xi_\mu\xi_5+m^5\xi_5.
}
The $4\times4$ $\xi$-matrices form a representation of the Clifford
algebra $\xi_\mu\xi_\nu+\xi_\nu\xi_\mu=2\delta_{\mu\nu}$, and are
identified with $SU(4)$ flavor generators
in the continuum limit.  We will denote
the terms in eq.\ \stagmass\
with scalar (S), vector (V), tensor (T), axial vector
(A) and
pseudoscalar (P) respectively.
This expression for $M$ can be derived from the fact that
each shift in the $\mu$-direction
of the field $\chi$, accompanied by a multiplication with $\zeta_\mu$,
corresponds to a multiplication of the continuum four-flavor Dirac field
$\psi$ by the matrix $\xi_\mu$:

\eqn\stagcont{
\zeta_\mu(x)\chi(x+{\hat\mu})\to\xi_\mu\psi(x).
}

It can be shown that this form of the mass matrix is stable under
renormalization, in the sense that the coefficients $m$, $m_\mu$,
$\dots$ will only receive multiplicative renormalizations, one for each
tensor structure in eq.\ \stagmass.  This was explicitly demonstrated
to one loop, and supplemented with more general symmetry arguments, in
ref.\ \refs{\golsmi}.  Due to the presence, in the massless theory, of
shift symmetries and
a continuous chiral symmetry (the so-called
$U(1)_\epsilon$ symmetry) there are no additive counterterms.  Note that
the mass matrix $M$ needs to be  diagonalized in order to
determine what the mass eigenstates are.

Let us first consider the simplest possible mass matrix, by choosing
only the single site mass $m$ in eq.\ \stagaction\ to be nonzero,
corresponding to four degenerate flavors.  In this case, the simplest
operator for an $\eta'_{SU(4)}$ will be

\eqn\stagetap{
\eqalign{
\eta'(x)\propto\;&\bar{\chi}(x)\epsilon(x)\zeta_1(x)\zeta_2(x+{\hat 1})
\zeta_3(x+{\hat 1}+{\hat 2})\zeta_4(x+{\hat 1}+{\hat 2}+{\hat 3})
\chi(x+{\hat 1}+{\hat 2}+{\hat 3}+{\hat 4})\cr
&+{\rm sum\ over\ all\
permutations\ on\ the\ directions\ }{\hat 1},\ {\hat 2},\ {\hat 3}\
{\rm and\ }{\hat 4},
}
}
which in the continuum limit corresponds to the operator
$\bar{\psi}\gamf\psi$ \refs{\golsmi},
where $\psi$ is a continuum Dirac field with four
flavor components.  In eq.\ \stagetap, the lattice gauge fields are
implicit.

In this basis, an $\eta'_{SU(2)}$ would be created by the continuum
operator

\eqn\conteta{
\eta'^{\rm cont}_{SU(2)}=\bar{\psi}
\pmatrix{1&0&0&0\cr 0&1&0&0\cr 0&0&0&0\cr 0&0&0&0\cr}\gamf\psi.
}
Clearly, in order to construct a staggered operator with this continuum
limit, we need the operator S to get a nonzero trace because the
$\eta'_{SU(2)}$ flavor matrix in
eq.\ \conteta\ has a nonvanishing trace, and V, T, A and P are all
traceless.  In addition,
we need an operator of the type V, T, A
and P, since the matrix contains
two zero eigenvalues.  The fact that these operators renormalize
differently from S leads to the need to tune their relative coefficient.
We conclude that with a single
site mass term no explicit $\eta'_{SU(2)}$
operator can be constructed in
the four-flavor staggered theory without tuning.  The only way to avoid
tuning in this case, is to compute the diagrams for the $\eta'_{SU(4)}$
(eq.\ \stagetap), and adjust the relative coefficients of the
straight-through and the two-hairpin diagrams as discussed at the end of
section 2.

Actually, the special properties of the tensor operator make it possible
to construct an $\eta'_{SU(2)}$ without tuning in a different way.
To discuss this, we will choose an explicit representation of the
$\xi$-matrices:

\eqn\ximatrices{
\xi_i=\sigma_i\otimes\tau_1,\ \ \
\xi_4=\tau_2,\ \ \
\xi_5=\tau_3.
}

In this case,
it is necessary to choose a mass term of the tensor type. For
definiteness we choose

\eqn\tensormass{
M_0=m(-i\xi_1\xi_2)=\pmatrix{m&0&0&0\cr 0&-m&0&0\cr
0&0&m&0\cr 0&0&0&-m\cr},
}
which corresponds again to four flavors with a degenerate mass $m$.
The minus signs can be removed by a nonanomalous chiral transformation.
The $\eta'_{SU(4)}$ with this mass matrix is

\eqn\contetafour{
\eta'^{\rm cont}_{SU(4)}\propto
\bar{\psi}(-i\xi_1\xi_2)\gamf\psi
=\bar{\psi}
\pmatrix{1&0&0&0\cr 0&-1&0&0\cr 0&0&1&0\cr 0&0&0&-1\cr}\gamf\psi\ .
}
Projecting to $SU(2)$, we get for the $\eta'_{SU(2)}$

\eqn\contetatwo{
\eta'^{\rm cont}_{SU(2)}\propto
\bar{\psi}
\pmatrix{1&0&0&0\cr 0&-1&0&0\cr 0&0&0&0\cr 0&0&0&0\cr}\gamf\psi
=\bar{\psi}(-i\xi_1\xi_2-i\xi_3\xi_4)\gamf\psi.
}
Unlike the previous case, this $\eta'_{SU(2)}$ flavor matrix is now
traceless, which allows us to write it as a sum of two tensor terms.
The minus sign which appears in this equation is removed by the same
chiral tranformation that removes the sign in the mass matrix,
eq.\ \tensormass.  The relevant observation here is that this $\eta'_{SU(2)}$
flavor matrix is not traceless with respect to the mass matrix, i.e.
$\tr\(M_0(-i\xi_1\xi_2-i\xi_3\xi_4)\)\ne 0$.  This follows from the fact
that the mass matrix defines what the flavor symmetries
are (in the continuum limit).  If $\psi_L$
and $\psi_R$ transform under $SU(4)_L\otimes SU(4)_R$ as

\eqn\psitransf{
\psi_L\to V_L\psi_L,\ \ \ \psi_R\to V_R\psi_R,
}
the condition for invariance is

\eqn\invcond{
V^\dagger_L M V_R = M.
}
With a degenerate mass matrix as in eq.\ \tensormass, the symmetry
group is $SU(4)$ (in the continuum limit).  If a chiral transformation
is performed to remove the minus signs in eq.\ \tensormass, eq.\ \invcond\ and
the trace condition
take on their usual form.

As mentioned above, the $\eta'_{SU(2)}$
of eq.\ \contetafour\ is now constructed from two tensor
operators
rather than one scalar and one of some other type.  Since all tensor
operators get renormalized in the same way, no tuning is needed here.
The price, however, is the use of a tensor mass term, which would
make this approach awkward for standard simulations.  Also, in the case of
a tensor mass, in general the staggered fermion determinant is not positive
(\cf\ the introduction to this section).
Using the
$\eta'_{SU(4)}$ and readjusting the relative weight of the diagrams by hand,
as explained in section 2, will
be preferable in most cases.

\newsec{Conclusion}

In this paper our investigations of ChPT in the quenched
approximation of QCD are extended
to theories in which only some of the quarks are
quenched.

The results are formulated in three theorems.  Two of them state
that in the subsector with unquenched valence quarks the theory is
equivalent to an unquenched theory with the number of flavors equal to
the number of unquenched quarks in the partially quenched theory.  The
super-$\eta'$ of the partially quenched theory is equivalent to the
$\eta'$ of this unquenched theory.

The third theorem deals with the existence of infrared divergences due
to the double pole in the quenched $\eta'$ twopoint function
\refs{\usPRD-\sharpetwo}.
Such divergences only arise if a particular quark mass scale is
completely quenched.  They do not show up in correlation functions with
only partially quenched or unquenched quarks on the external lines.

Some  one-loop calculations serve as explicit examples of these results.
Moreover, we apply the $n=4$, $k=2$ case
to staggered fermion QCD with a single
site mass term, in which the square root of the fermion determinant is
taken in order to yield two-flavor QCD.  Our analysis shows that this
technique is valid within ChPT, and that the super-$\eta'$ of the
$SU(4|2)$ theory (or equivalently the $SU(4)$ $\eta'$ with by-hand
reweighting of diagrams) can be used to measure
$SU(2)$ $\eta'$ correlation functions without any of the fine tuning
which is often necessary for staggered fermions.

Finally, we have shown that another two-flavor $\eta'$ operator,
not based on the super-$\eta'$,  exists
for which no fine tuning is needed if one employs  staggered fermions
with a so-called tensor mass.

\bigskip
{\bf Acknowledgements}

We are greatful to Doug Toussaint for inspiring this
project by suggesting that our chiral techniques
could be used to examine the determinant square-root method
for staggered fermions.  We also thank Mike Ogilvie for very
useful discussions, and Carl Bender for the proof of the lemma
presented in the appendix.
Part of this work was carried out at Los Alamos National Laboratory and
UC Santa Barbara.  M.G. would
like to thank Rajan Gupta and the Theory Division of LANL,
and both of us would like to thank the UCSB Physics Department,
and in particular Bob Sugar, for hospitality.

C.B. and M.G. are supported in part by the Department of Energy under grant
\doe.

\vfill\eject
{\bf Appendix}

In this appendix we present a proof\foot{This proof is due to C. Bender.}
 of the lemma that we
used in section 3.  The lemma states that the function

\eqn\fz{
f(z)=1+\sum_{i=1}^k\bigfrac{1}{z-\alpha_i}
}
has no double zeros if all the $\alpha_i$ are real.  The proof will
be by contradiction.
So let us assume that $f(z)$ has a double zero
at $z=\beta$.  First, $f(z)$ diverges when $z$ is equal to any
of the $\alpha_i$, so we can assume that $\beta\ne\alpha_i$ for all $i$.
Now define  $z'=z-\beta$.  Then $f(z')$, which is given by

\eqn\fzprime{
f(z')=1+\sum_{i=1}^k\bigfrac{1}{z'-\alpha'_i}
}
with $\alpha'_i=\alpha_i-\beta$, now has a double zero at $z'=0$,
with all $\alpha_i\ne 0$.
Note that the $\alpha'_i$ are not necessarily real, but can have a
common imaginary part.  From now on we will drop the primes on $z$
and  $\alpha_i$ and assume that $f(z)$ has a double zero at $z=0$.
$f$ can be written as

\eqn\faspol{
f(z)=\bigfrac{P_k(z)}{\prod_{i=1}^k(z-\alpha_i)},
}
where $P_k(z)$ is a polynomial in $z$ of degree $k$:

\eqn\pol{
P(z)=z^k+\dots+\prod_{i=1}^k(-\alpha_i)\left[-\sum_{i=1}^k
\bigfrac{1}{\alpha_i}+\sum_{i\ne j}\bigfrac{1}{\alpha_i\alpha_j}\right]z
+\prod_{i=1}^k(-\alpha_i)\left[1-\sum_{i=1}^k\bigfrac{1}{\alpha_i}\right].
}

If $f(z)$ has a double zero at $z=0$, the constant term and the coefficient
of the linear term in $P(z)$ have to vanish, \ie,

\eqn\coeff{\sum_{i=1}^k\bigfrac{1}{\alpha_i}=1,\qquad
\sum_{i=1}^k\bigfrac{1}{\alpha_i}=
\sum_{i\ne j}\bigfrac{1}{\alpha_i\alpha_j}=\left(\sum_{i=1}^k
\bigfrac{1}{\alpha_i}\right)^2-\sum_{i=1}^k\bigfrac{1}{\alpha_i^2}.
}
 From the first of these equations one concludes that the common imaginary
part of the $\alpha_i$ has to vanish.  Therefore the $\alpha_i$ have to
be real (in other words, the original
double zero would have to be on the real axis).
Substituting the first equation into the second, we then conclude that

\eqn\sumsq{
\sum_{i=1}^k\bigfrac{1}{\alpha_i^2}=0,
}
which has no solution for real $\alpha_i$.
This completes the proof that the function $f(z)$
has no double zeros anywhere in the complex plane.

\listrefs

\vfill
\bye